\documentclass[showpacs,twocolumn,superscriptaddress,nofootinbib]{revtex4-1} 

\usepackage[utf8]{inputenc}
\usepackage{hyperref}
\hypersetup{colorlinks=true,linkcolor=blue,citecolor=cyan}
\usepackage{graphicx}
\usepackage{xcolor, soul}
\sethlcolor{green}
\usepackage{dcolumn}
\usepackage{bm}
\usepackage{color}
\usepackage{enumitem}
\usepackage{mathrsfs}
\usepackage{amsmath}
\usepackage{amssymb}
\usepackage{cleveref}

\usepackage{soul,xcolor}
\setstcolor{red}

\begin{document}

\title{Magnetic reconnection in five-dimensional Kerr black hole}

\author{Ikhtiyor Eshtursunov}
\email{eshtursunovikhtiyor@gmail.com}
\affiliation{New Uzbekistan University, Movarounnahr str. 1, Tashkent 100000, Uzbekistan}

\author{Sanjar Shaymatov}
\email{sanjar@astrin.uz}
\affiliation{Institute of Fundamental and Applied Research, National Research University TIIAME, Kori Niyoziy 39, Tashkent 100000, Uzbekistan}
\affiliation{University of Tashkent for Applied Sciences, Str. Gavhar 1, Tashkent 100149, Uzbekistan}
\affiliation{National University of Uzbekistan, Tashkent 100174, Uzbekistan}

\author{Chengxun Yuan}
\email{yuancx@hit.edu.cn}
\affiliation{School of Physics, Harbin Institute of Technology, Harbin 150001, People’s Republic of China}

\date{\today}
\begin{abstract}

In this paper, we employ the Comisso–Asenjo magnetic reconnection (MR) mechanism to investigate energy extraction from a rapidly rotating five-dimensional Kerr black hole (BH) with single- and two-rotation configurations. We analyze the efficiency, phase-space structure of accelerated and decelerated plasma energies, and the extracted power as functions of the spin parameter, reconnection location, plasma magnetization, and magnetic field orientation. We show that MR significantly enhances energy extraction from a five-dimensional BH with a single rotation and that the extraction efficiency is higher in the single rotation configuration than in the two-rotation case. We also evaluate the extraction rate and compare it with the Blandford–Znajek (BZ) mechanism, showing that the extracted power can exceed that of the BZ process in the single-rotation configuration. Our analysis shows that MR can significantly improve energy extraction in five-dimensional Kerr BHs with a single rotation, making them promising candidates for powering high-energy astrophysical phenomena.

\end{abstract}
\pacs{}
\maketitle

\section{Introduction}
\label{introduction}

Black holes (BHs), formed through the gravitational collapse of massive stars, are a generic prediction of general relativity (GR). Their existence is strongly supported by gravitational-wave (GW) detections from BH mergers \cite{Abbott16a,Abbott16b} and direct imaging by the Event Horizon Telescope (EHT) \cite{Akiyama19L1,Akiyama19L6}. These observations provide compelling evidence for BHs and enable tests of strong-field gravity near the horizon. Astrophysical BHs are widely regarded as the central engines of highly energetic phenomena, including active galactic nuclei, X-ray sources, and gamma-ray bursts \cite{King01ApJ,Peterson:97book,Meszaros06}, with luminosities reaching $10^{42}-10^{47}\:\rm{erg/s}$ \cite{Fender04mnrs,Auchettl17ApJ,IceCube17b}. Understanding these processes requires efficient energy extraction mechanisms for extracting rotational energy from rapidly rotating BHs. 

Energy extraction from rotating BHs, first demonstrated by Penrose via particle splitting in the ergosphere, is a fundamental prediction of general relativity \cite{Penrose:1969pc}. This Penrose process (PP) extracts rotational energy from a BH in the ergosphere, where negative-energy states are allowed. While one fragment falls into the BH, the other escapes with enhanced energy. However, the process is limited by its low efficiency and small cross section. Hence, this mechanism has limited relevance for understanding BH–driven high energy phenomena. Later, the magnetic Penrose process (MPP), extends the original PP by allowing a weak magnetic field to supply the energy needed for particles to access negative-energy orbits, overcoming its velocity constraints \cite{Bhat85,Parthasarathy86ApJ}. Extensions of the original PP, including the MPP and the Blandford–Znajek (BZ) mechanism \cite{Blandford1977}, are widely used to explain high-energy astrophysical phenomena \cite{Bhat85,Parthasarathy86ApJ,Blandford1977,Rees:1984si,Peterson:97book,Meszaros06,King01ApJ,McKinney07}. While the BZ process relies on magnetohydrodynamic interactions of the force-free plasma, the MPP extracts energy via frame dragging and electromagnetic effects and can achieve efficiencies exceeding 100\% for charged particles \cite{Abdujabbarov11,Dhurandhar1984PRD,Dhurandhar1984PRD.30.1625,Wagh85ApJ,Tursunov:2019oiq,Shaymatov24PRD.110d4042S,Xamidov24EPJC,Dadhich18MNRAS,Shaymatov24EPJC,Shaymatov22b,Khamidov25JCAP...03..053X}.

In addition to the mechanisms discussed above, other processes such as magnetic reconnection (MR) \cite{Koide08ApJ,Parfrey19PRL} and superradiance \cite{Brito20Book} have been proposed as viable mechanisms for the extraction of energy from BHs through high-energy particle acceleration, leading to extensive investigations in various astrophysical scenarios. In the MR mechanism, frame dragging near the BH horizon produces antiparallel magnetic field configurations, enabling this mechanism in the ergosphere. As in the MPP, this allows negative-energy particles to be absorbed by the BH while others escape with enhanced energy, resulting in rotational energy extraction. Slow magnetic reconnection is generally inefficient for powering the most energetic sources \cite{Koide08ApJ}. Efficient operation requires relativistic relative velocities between the resulting particle fragments, implying that relativistic MR is necessary to accelerate particles rapidly and generate energetic outflows \cite{Parfrey19PRL}. Therefore, MR needs to accelerate the particles to relativistic energies, generating powerful jets from BH \cite{Blandford:1982di,Marscher:2008aa,Kotera2011ARA&A,Romero:2008zj} and making this mechanism a strong candidate for high-energy astrophysical phenomena. However, a quantitative estimate of the extracted energy has yet to be determined. Later, Comisso and Asenjo \cite{Comisso21} proposed a MR-driven energy extraction mechanism for the rapidly rotating Kerr BH, allowing a quantitative evaluation of efficiency and power, which strongly depend on the spin of BH. The Comisso-Asenjo MR mechanism has been widely studied in recent works \cite{Liu22ApJ,Wei22,Carleo22,Khodadi22,Wang22,Khodadi23MR_JCAP,Shaymatov24MR}. Subsequent studies have applied this mechanism to rapidly rotating BHs, providing valuable formation on the energy extraction driven by MR relative to other mechanisms \cite{Zhang24JCAP...07..042Z,Chen24PRD.110f3003C,Shen25PRD.111b3003S,Long25EPJC...85...26L,Rodriguez25PDU,YuChih25PRD.112j4016Y,Wang25JCAP,Zeng25PRD.112f4032Z,Zeng25PRD.112f4080Z,Cheng25EPJC...85.1130C,Eshtursunov:2026kgr}.

BHs are among the most intriguing predictions of GR. In higher dimensions ($D>4$), rotating solutions such as the Myers–Perry (MP) BH \cite{Myers-Perry86} admit multiple rotation axes, leading to two independent spin parameters in five dimensions. This further motivates an investigation of their energetics. Exact charged solutions in five-dimensional Einstein–Maxwell supergravity have also been constructed \cite{Chong05a}. The energetic properties and energy-extraction processes of such higher-dimensional black holes with their dynamics have been studied in various works \cite{Nozawa05,Prabhu10,Abdujabbarov13bsw,Dadhich22a,Shaymatov24PRD.110d4042S} from different perspective. typically in the absence and presence of external magnetic fields. In addition, the linear and non-linear accretion processes around five-dimensional MP BHs have been extensively analyzed \cite{An18,Shaymatov19a,Shaymatov20a,Shaymatov20b,2022IJMPD..3150120D,2021PDU....3100758S}. In this work, we consider the five-dimensional MP rotating BH spacetime \cite{Myers-Perry86} and apply the Comisso–Asenjo mechanism to investigate energy extraction driven by MR for both single and two-rotation configurations. We find that energy extraction is more efficient in the single-rotation configuration than in both the two-rotation configuration and the four-dimensional Kerr BH. The areas of phase-space for the energies of accelerated and decelerated plasma and the extracted power are further analyzed as a function of spin, reconnection location, plasma magnetization, and magnetic field orientation. 

The paper is organized as follows: In Sec.~\ref{Sec:MP}, we briefly introduce the five-dimensional Kerr (MP) BH spacetime and discuss its relevant properties. In Sec.~\ref{Sec:MP_MR}, we apply the Comisso–Asenjo mechanism to investigate energy extraction via MR process around a rotating Kerr BH. We then analyze the efficiency, phase-space regions for the energies of accelerated and decelerated plasma and power of energy extraction, as well as the overall energy extraction rate, with the results presented in Sec.~\ref{Sec:MP-Power-MR}. Finally, we summarize our findings in Sec.~\ref{Sec:con}. Throughout this work, we employ a system of geometric units with $G_{\rm{N}}=c=1$.

\section{Five dimensional rotating Kerr (Myers-Perry) black hole}\label{Sec:MP}
 
It is known \cite{Comisso21} that magnetic reconnection in the ergoregion, driven by frame dragging that twists magnetic field lines around a rapidly rotating BH, provides an efficient mechanism for extracting rotational energy. Combining this mechanism with our results, we find that energy extraction is more efficient for a five-dimensional rotating BH with a single rotation parameter than for configurations with two rotations, as well as compared to the four-dimensional Kerr BH. We begin with the metric describing a five dimensional Myers-Perry (MP) rotating BH \cite{Myers-Perry86,Hawking99},
\begin{eqnarray}\label{5D_metric}
ds^2&=&-\frac{\Delta}{\Sigma}\left(dt-a\sin^2\theta
d\phi-b\cos^2\theta
d\psi\right)^2+\frac{\Sigma}{\Delta}dr^{2}\nonumber\\
&+&\Sigma d\theta^2 +
\frac{\sin^2\theta}{\Sigma}\left[(r^2+a^2)d\phi-a
dt\right]^2\nonumber\\&+&\frac{\cos^2\theta}{\Sigma}\left[(r^2+b^2)d\psi-b
dt\right]^2 \nonumber\\&+&
r^2\left(\cos^2\theta+\sin^2\phi\right)d\psi^2\ ,\ \quad
\end{eqnarray}
with 
$$
\Delta=\frac{(r^{2}+a^{2})(r^{2}+b^{2})}{r^{2}}-\mu\, ,
$$ and
$$\Sigma=r^{2}+a^{2}\cos^{2}\theta+b^{2}\sin^{2}\theta\, .$$ 
We note that {${\mu=\frac{8M}{3\pi}}$} is the mass parameter, while $a=\frac{4J_{\phi}}{\pi\mu}$
and $b=\frac{4J_{\psi}}{\pi\mu}$ are the spin parameters associated with two rotation axes. The horizon equation $\Delta=0$ solves to give 
 \begin{eqnarray}
 r_{\pm}&=&\left(\frac{1}{2}\right)^{1/2}\left[\left({\mu}-a^{2}-b^{2}\right)\right.\nonumber\\
 &\pm & \left.\sqrt{\left({\mu}-a^{2}-b^{2}\right)^{2}-4a^2b^{2}} \right]^{1/2}\, ,
 \end{eqnarray}
where $a + b = (\mu)^{1/2}$ refers to an extremality condition. 
 
It is worth noting that, due to stationarity and axial symmetry, the five-dimensional MP rotating BH admits three Killing vectors \cite{Myers-Perry86}, i.e. 
\begin{eqnarray}
    \xi_{(t)} = \frac{\partial}{\partial t} \ , \quad \xi_{(\phi)}= \frac{\partial}{\partial \phi} \quad \mbox{and} \quad \xi_{(\psi)}=\frac{\partial}{\partial \psi} \, .
\end{eqnarray} 
Here, $\xi_{(t)}$ corresponds to stationarity, while $\xi_{(\phi)}$ and $\xi_{(\psi)}$ generate axial symmetries in five-dimensional spacetime. These Killing vectors imply three conserved quantities for a test particle of mass m: the energy and two independent angular momenta. They also provide the basis for constructing the electromagnetic vector potential of a Maxwell test field in the five-dimensional MP spacetime. It should be emphasized that the n-velocity of a zero-angular-momentum observer (ZAMO) in the five dimensional MP rotating BH spacetime reads as \cite{Shaymatov24PRD.110d4042S} 
 \begin{align} 
 u^\mu = \alpha\left(\xi_{(t)}^\mu +\Omega_{\phi} \xi_{(\phi)}^\mu + \Omega_{\psi} \xi_{(\psi)}^\mu\right)\, . 
 \end{align}
We note that $u^\mu$ is orthogonal to the hypersurfaces of the constant $t = const$, implying $u^r=0$ and $u^{\theta_{i}}=0$. The normalization factor $\alpha$ is determined from the condition $g_{\mu\nu}u^\mu u^\nu =-1$. We now consider the static limit surface, defined by the vanishing of the norm of the time-like Killing vector $\xi_{(t)}$, i.e., $g_{tt}=0$, which implicitly determines $r_{st}$ as \cite{Prabhu10}: 
\begin{eqnarray}
    r_{st} = \sqrt{\mu-a^2\cos^2\theta -b^2\sin^2\theta}\, .
\end{eqnarray}
Energy extraction through magnetic reconnection requires the presence of an ergosphere, defined by $r_{+} < r < r_{st}$, where $r_{+}$ is the horizon and $r_{st}$ the static limit. In five dimensions, the static limit surface differs from that of the four-dimensional Kerr black hole, leading to a modified ergosphere structure. Motivated by this, we examine magnetic reconnection in the vicinity of the five-dimensional MP BH.

\section{MAGNETIC RECONNECTION AS A MECHANISM OF ENERGY EXTRACTION}\label{Sec:MP_MR}

In our universe, BHs are considered one of the most energetic and powerful candidates of astrophysical compact objects; therefore, their energetics are always the theme of scientific investigations. Going more deeply into the energetic properties of five-dimensional BHs is an essential key to understanding more about the high-energy astrophysical phenomena and its important aspects. Hence, here we focus on the energy extraction from five-dimensional Kerr BH by using the Comisso-Asenjo MR mechanism~\cite{Comisso21} and examine plasma magnetization as a result of frame dragging effects in the magnetic field lines around the BH. For our investigation, it is convenient to use a frame that is locally nonrotating and is called ZAMO, in which we see plasma energetic properties including energy density at infinity. Our work considers that for bulk plasma, the reconnection occurs in the equatorial plane of the BH (i.e., $\theta=\pi/2$). We should note that in the ZAMO frame the spacetime line element reads as: 
  \begin{eqnarray}
      ds^2=-d\hat{t}^2+\sum_{i=1}^k(d\hat{x^i})^2=\eta_{\mu \nu}\hat{dx^{\mu}}\hat{dx^{\nu}}\, . 
  \end{eqnarray}
The Boyer-Lindquist coordinates and the ZAMO coordinates can be transformed from one to another as
\begin{eqnarray}
d\hat{t}=\alpha dt \mbox{~~and~~} d\hat{x^i}=\sqrt{g_{ii}}dx^i-\alpha \beta^{i}dt\, ,
\end{eqnarray}
with $\alpha$ indicating the lapse function and with {$(\beta^r, \beta^\theta, \beta^\phi, \beta^\psi)$ $\to$ $(0,0,\beta^\phi,0)$ referring to the shift vector in the equatorial plane of a five-dimensional Kerr BH}, which are given by
\begin{eqnarray}
\alpha=\sqrt{-g_{tt}+\frac{g_{\phi t}^2}{g_{\phi \phi}}}  \mbox{~~and~~}\beta^{\phi}=\frac{\sqrt{g_{\phi\phi}} \omega^{\phi}}{\alpha}\, ,
\end{eqnarray}
where $\omega^{\phi}=-{g_{\phi t }}/{g_{\phi \phi}}$ is the frame dragging angular velocity.
\begin{figure} 
\begin{center} \begin{tabular}{c c} 
\includegraphics[scale=0.65]{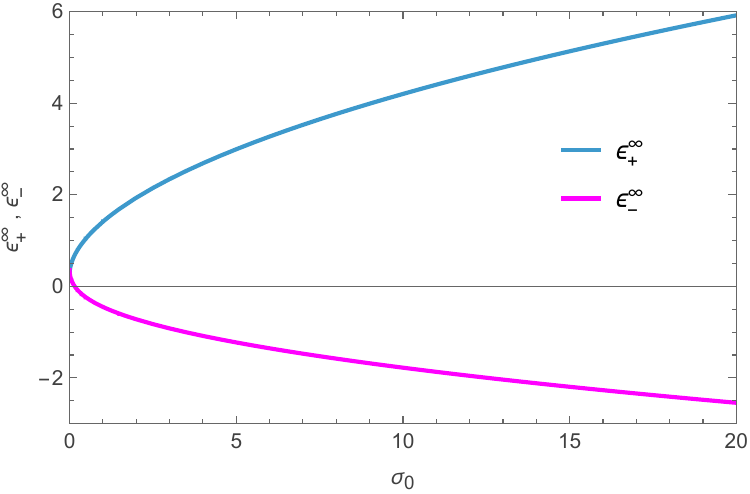}\hspace{1cm} 
\end{tabular} \caption{\label{Fig:energyatinf} {The relation between accelerated and decelerated plasma energies per enthalpy at infinity and magnetization parameter is plotted when $\xi$ $\to $ 0 and a = b $\to$ 0.5. }} 
\end{center}
\end{figure}
\begin{figure*} 
\begin{center} \begin{tabular}{c c} 
\includegraphics[scale=0.65]{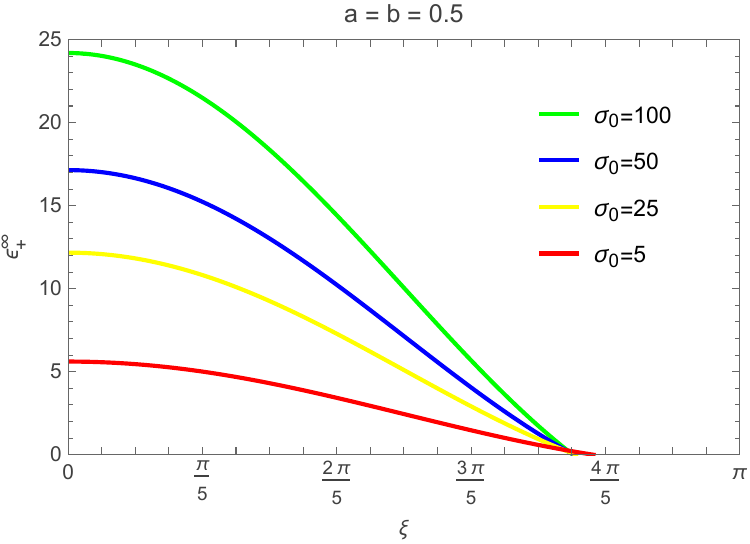}\hspace{1cm} 
\includegraphics[scale=0.65]{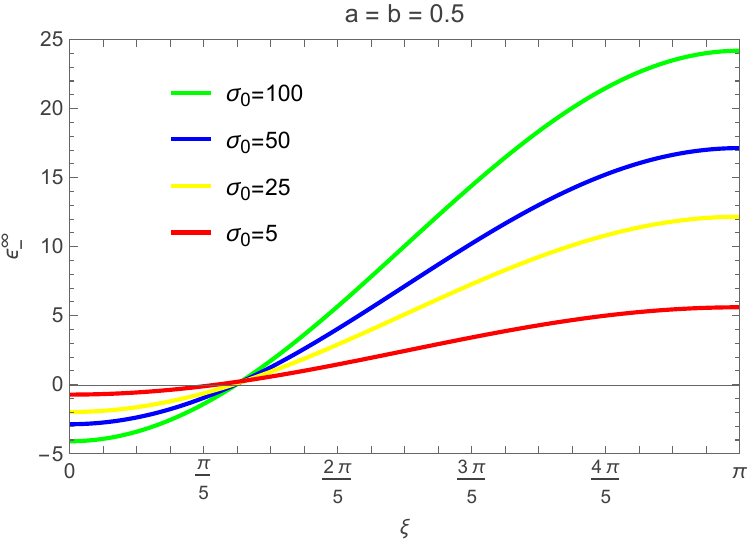}\hspace{1cm}
\end{tabular} \caption{\label{Fig:energyangle} {The behavior of energies at infinity per enthalpy of accelerated and decelerated plasma against the orientation angle of reconnecting magnetic field around the five-dimensional Kerr BH with two-rotation configuration, $a=b=0.5$. The energy curves vary (red to green) with the corresponding values of magnetization parameters (from $\sigma_0$=5 to $\sigma_0$=100). }} 
\end{center}
\end{figure*}
\begin{figure*} 
\begin{center} \begin{tabular}{c c} 
\includegraphics[scale=0.65]{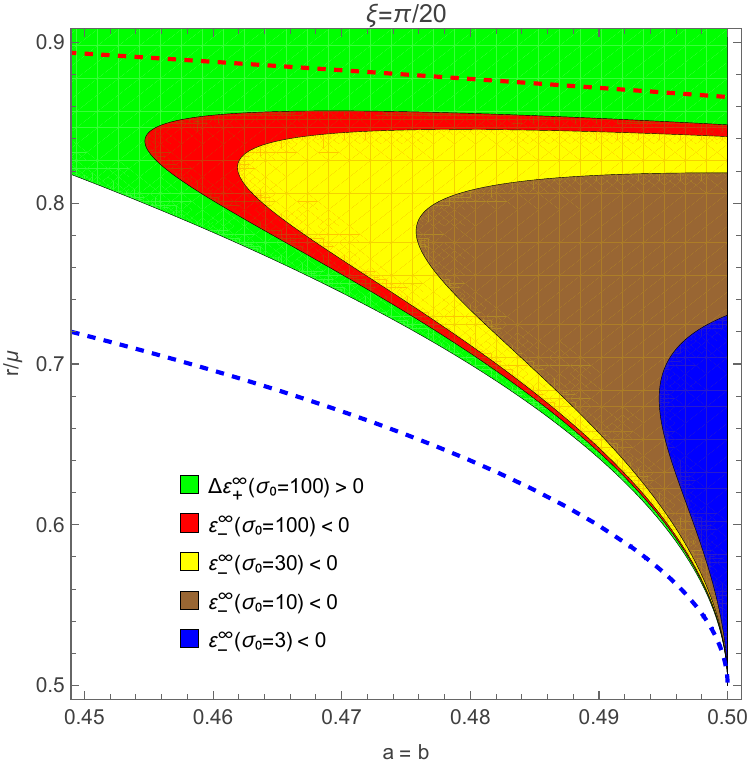}\hspace{1cm} 
\includegraphics[scale=0.65]{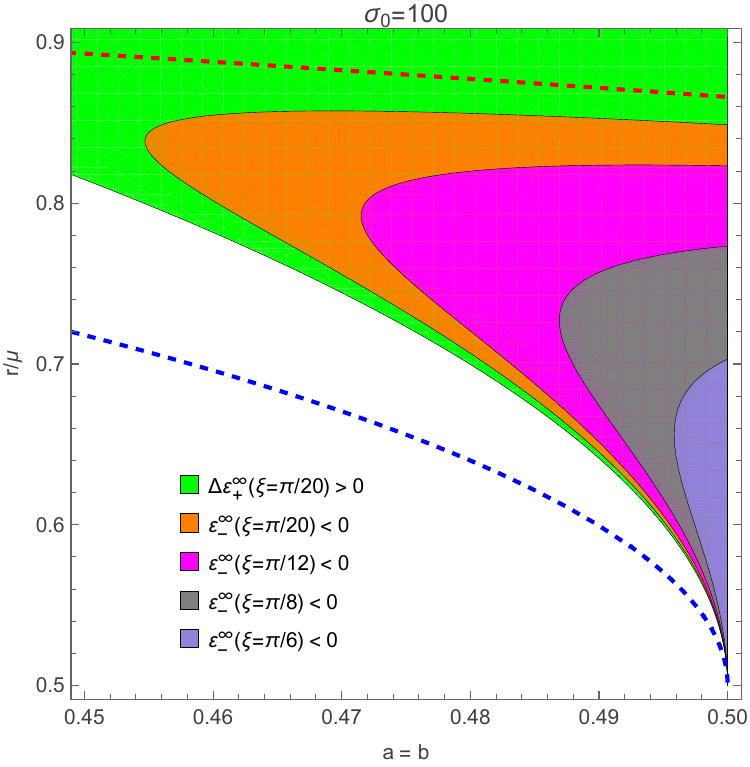}\hspace{1cm}
\end{tabular} \caption{\label{Fig:energyregion} {The areas of phase-space for the energies of accelerated ($ \epsilon_{+}^{\infty} > 0$) and decelerated ($\epsilon_{-}^{\infty} < 0$) plasma. Left: The allowed regions are shown for different values of magnetization parameter (from $\sigma_0$=3 to $\sigma_0$=100) while the orientation angle $\xi$=$\pi/20$. Right: The allowed regions are shown for various combinations of orientation angle (from $\xi$=$\pi/20$ to $\xi$=$\pi/6$) when the magnetization is $\sigma_0$=100. Dashed blue/red lines represent the outer event horizon and ergosphere's outer boundary, respectively.}}
\end{center}
\end{figure*}
\begin{figure*}
    \centering
    \includegraphics[scale=0.7]{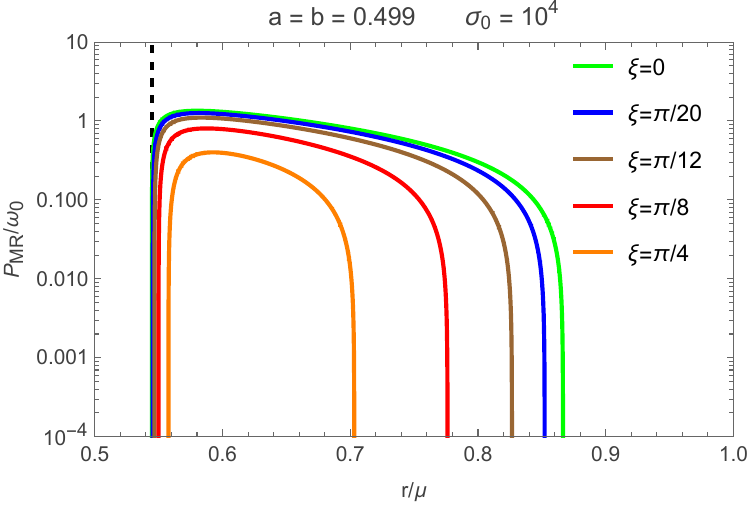}
    \includegraphics[scale=0.7]{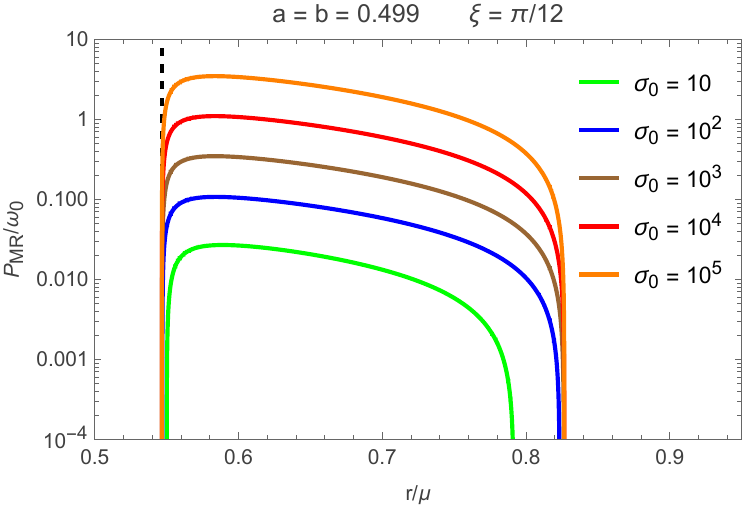}
    \caption{\label{fig:efficiency} {The power extracted from five-dimensional Kerr BH by using MR is shown when rotation parameters maximally equal $a=b$. Left panel shows the relation between the power and the location $r/\mu$ for different quantities of orientation angle from $\xi$=0 to $\xi$=$\pi/4$ (green to orange lines) and fixed value of magnetization parameter $\sigma=10^4$. Right panel represents the connection for various values of magnetization parameter from $\sigma_0$=10 to $\sigma_0$=$10^5$ (green to orange lines) and fixed orientation angle $\xi$=$\pi/12$. }}
\end{figure*}

We assume that the plasma co-rotates in the equatorial plane $\theta=\pi/2$. To examine its dynamics, we then write the corresponding Keplerian angular velocity as follows:
\begin{eqnarray}
\Omega_K=\frac{d\phi}{dt}=\frac{-\partial_r g_{t\phi}\pm\sqrt{(\partial_r g_{t\phi})^2-(\partial_r g_{tt})(\partial_r g_{\phi\phi})}}{\partial_r g_{\phi\phi}}\, .
\end{eqnarray}
For any vector $h$, its covariant and contravariant components, when transformed to the ZAMO frame, can be expressed as: $\hat{h_0}=\frac{h_0}{\alpha}+\sum_{i=1}^3 \frac{\beta^i}{g_{ii}}h_i \mbox{~~and~~} \hat{h_i}=\frac{h_i}{\sqrt{g_{ii}}}\, ; \mbox{~~}
   \hat{h^0}=\alpha h^0 \mbox{~~and~~} \hat{h^i}=\sqrt{g_{ii}}h^i-\alpha \beta^i h^0\, .$
By using the transformations, in the frame of ZAMO corotating plasma's Keplerian velocity is expressed as
\begin{eqnarray}
 \hat{v}_K&=&\frac{d\hat{x}^{\phi}}{d\hat{t}}
 =\frac{\sqrt{g_{\phi \phi}}}{\alpha}\Omega_K-\beta^{\phi}\, . 
 \end{eqnarray}
In the one-fluid (magnetohydrodynamic) approximation, the plasma energy-momentum tensor takes the form of 
\begin{eqnarray}
    T^{\mu \nu}=pg^{\mu \nu}+{\mathit{w}}U^{\mu}U^{\nu}+F^{\mu}_{\delta}F^{\nu \delta}-\frac{1}{4}g^{\mu \nu}F^{\rho \delta}F_{\rho \delta}\, , 
\end{eqnarray}
in the above equation $p$ defines plasma pressure and $\omega$ and $U^{\mu}$ refer to enthalpy density and the four-velocity, while $F^{\mu\nu}$ to the electromagnetic field tensor.

MR takes place in the ergoregion of the BH and is much more similar to the PP case \cite{Penrose:1969pc}. There are two plasmas investigated which are accelerated one gains BH's positive energy and the second one - decelerates and by taking the negative energy that falls into the BH.  
 \begin{eqnarray}
e^{\infty}=-\alpha g_{\mu 0}T^{\mu 0}=\alpha \hat{e}+\alpha \beta^{\phi}\hat{P}^{\phi}\, .
\end{eqnarray}
In the above equation, $\hat{e}$ is the density of total energy and $\hat{P}^{\phi}$ refers to the momentum density's azimuthal component, and they are introduced as
\begin{eqnarray}
    \hat{e}={\mathit{w}}\hat{\gamma}^2-p+\frac{\hat{B}^2+\hat{E}^2}{2}\, ,\\
   \hat{P}^{\phi}={\mathit{w}}\hat{\gamma}^2\hat{v}^{\phi}+(\hat{B} \times \hat{E})^{\phi}\, ,
\end{eqnarray}
$\hat{v}^{\phi}$ refers to the plasma velocity's $\phi$-component at the ZAMO and $\hat{\gamma}$ is the Lorentz factor $\hat{\gamma}=\hat{U}^0=({1-\sum_{i=1}^3(d\hat{v}^i)^2})^{-1/2}$. Electric field component $\hat{E}^i$ is defined as
\begin{eqnarray}
 \hat{E}^i=\eta^{ij}\hat{F}_{j0}=\hat{F}_{i0}\, .
\end{eqnarray}
The magnetic field $\hat{B}^i$ can be expressed as
\begin{eqnarray}
\hat{B}^i=\epsilon^{ijk}\hat{F}_{jk}/2\, .  
\end{eqnarray}
The electromagnetic and hydrodynamic components add up to the energy density at infinity $e^{\infty}=e_{hyd}^{\infty}+e_{em}^{\infty}$. In this instance, we examine plasma as incompressible, adiabatic matter, hence, according to the Comisso-Asenjo mechanism~\cite{Comisso21}, the electromagnetic energy density contribution to the overall energy density is thought to be extremely low. As a result, the hydrodynamic energy density at infinity in the ZAMO system can be expressed as the total energy density at infinity:
\begin{eqnarray}
 e^{\infty}\approx e_{hyd}^{\infty}=\alpha \hat{e}_{hyd}+\alpha \beta^{\phi}{\mathit{w}}\hat{\gamma}^2\hat{v}^{\phi}\
\end{eqnarray}
In the ZAMO, taking into account the energy density of the hydrodynamic field $\hat{e}_{hyd}={\mathit{w}}\hat{\gamma}^2-p$ and we can express the energy density at infinity as follows:
   \begin{equation}
    e^{\infty}=e^{\infty}_{hyd}=\alpha [{\mathit{w}}\hat{\gamma}(1+\beta^{\phi}\hat{v}^{\phi})-\frac{ p}{\hat{\gamma}}]\, .
\end{equation}
The magnetization of the plasma upstream relative to the reconnection layer is related to the azimuthal component of the plasma outflow velocity and the corresponding Lorentz coefficient in the Comisso-Asenjo MR mechanism as~\cite{Comisso21}
\begin{equation}
v_{out}\approx (\frac{\sigma_0}{1+\sigma_0})^{1/2} ,
\end{equation}
\begin{equation}
\gamma_{out}= (1-v_{out}^2)^{-1/2}\approx (1+\sigma_0)^{1/2},
\end{equation}
where $B_0$ and $\sigma_0=B_0^2/\omega_0$ represent the magnetic field in large scales (i.e., the magnetization parameter), $\omega_0$ refers to the plasma enthalpy density, as we stated above. We then write the azimuthal components of the corotating and counterrotating plasma flow velocities as follows:
\begin{equation}
v_{\pm}^{\phi}=\frac{\hat{v}_K \pm v_{out}\cos{\xi}}{1 \pm \hat{v}_K v_{out}\cos{\xi}}\, ,
\end{equation}
in the above equation, $\xi$ is defined as the angle between the radial and azimuthal velocity components in the outflow, i.e., it is referred to as the orientation angle of the magnetic field where the magnetic reconnection occurs. By combining the equations above, the form of the energy density at infinity per enthalpy $e^{\infty}/\omega$ can be expressed
\begin{eqnarray}\label{Eq:essential}
 \epsilon^{\infty}_{\pm}&=&\alpha \hat{\gamma}_K \Bigg[(1+\beta^{\phi}\hat{v}_K)\sqrt{1+\sigma_0}\pm \cos{\xi}(\beta^{\phi}+\hat{v}_K)\sqrt{\sigma_0}\nonumber\\
 &-&\frac{\sqrt{1+\sigma_0}\mp \cos{\xi}\hat{v}_K\sqrt{\sigma_0}}{4\hat{\gamma}^2(1+\sigma_0-\cos^2{\xi}\hat{v}_K^2\sigma_0)}\Bigg].  
\end{eqnarray}
It should be emphasized that the plasma considered in our investigation is taken as relativistically hot, i.e., $\omega=4p$. Negative energy-at-infinity defines the energetic state of decelerated plasma, and its sign is minus in small orientation angles, as shown in Fig.~\ref{Fig:energyangle}, whereas the extracted energy via MR mechanism has to be positive in order to harness BH energy:
\begin{eqnarray}
 \epsilon_{-}^{\infty}<0\, , 
 \end{eqnarray}
 \begin{eqnarray}
 \Delta \epsilon_{+}^{\infty}=\epsilon_{+}-\left(1-\frac{\Gamma}{\Gamma-1}\frac{p}{\mathit{w}}\right)>0\, .
\end{eqnarray}
in the above equation, for $\Gamma$, the value (4/3) is taken to express the relativistic hot plasma state.

We further assess the usage of energy extraction through the Comisso-Asenjo MR mechanism from five-dimensional Kerr BH. As written in Eq.~(\ref{Eq:essential}), the energies at infinity of decelerated and accelerated plasma per enthalpy have very complicated and long forms. Therefore we analyze relations of these energies with essential reconnection parameters as magnetic field orientation angle where reconnection happens and magnetization parameter. Figure ~\ref{Fig:energyatinf} indicates magnetization-energies at infinity connection for maximal possible energy extraction when $\xi$ $\to$ 0 and a = b $\to$ 0.5. As the parameter $\sigma_0$ increases, the energies' values rise ($\Delta \epsilon_{+}^{\infty}$) and decrease ($\epsilon_{-}^{\infty}$) monotonically and we should take this strong dependence into account to test energy extraction efficiency of the mechanism. In Fig.~\ref{Fig:energyangle}, we can see the energy dependence on the orientation angle for various magnetization parameters when rotation parameters maximally equal $a=b=0.5$. As we stated earlier, accelerated plasma energy per enthalpy at infinity should be positive and the decelerated one's should be negative for efficient energy extraction via MR. Therefore we should take orientation angle as $\xi<\pi/6$. Smaller orientation parameters lead to more efficient energy extraction.
\begin{figure*}
\centering
    \includegraphics[scale=0.68]{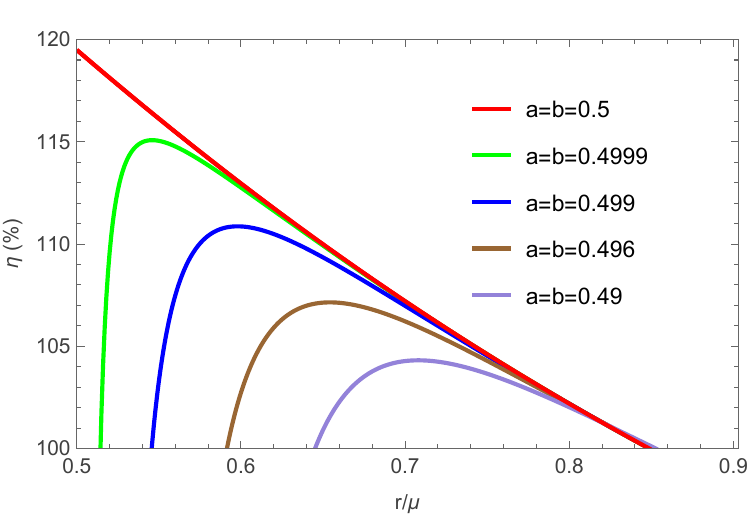}
    \includegraphics[scale=0.55]{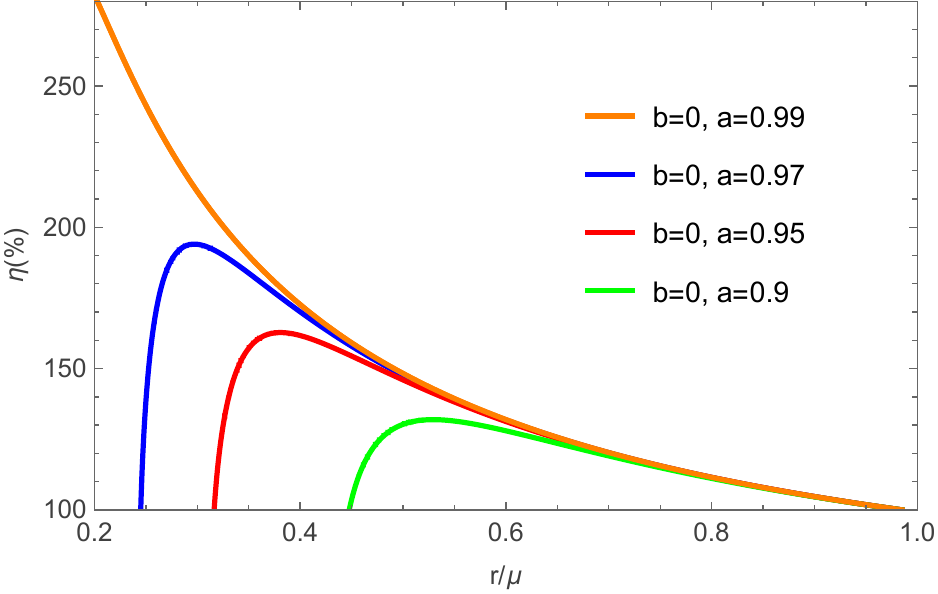}
    \caption{\label{fig:efficiencyeta1} {The efficiencies of energy extraction via the MR mechanism from a five-dimensional Kerr BH are plotted as a function of the point location $r/\mu$. Left panel: the two rotation parameters are equal and taken nearly maximal values. Right panel: one of the rotation parameters is zero, while the other is assumed to be a nearly maximum value. In both cases, we set $\sigma_0=100$ and $\xi=\pi/20$.}}
\end{figure*}
\begin{figure*}
\centering
    \includegraphics[scale=0.48]{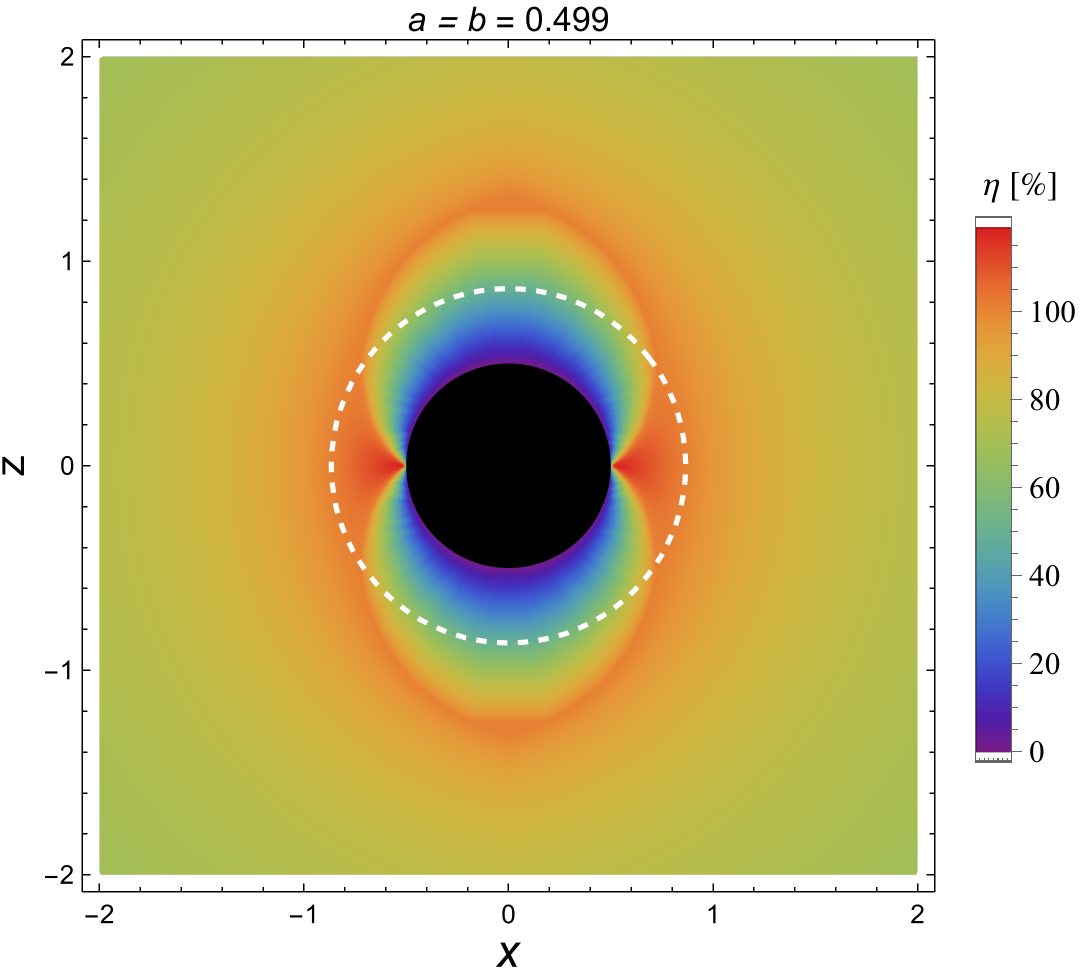}
    \includegraphics[scale=0.48]{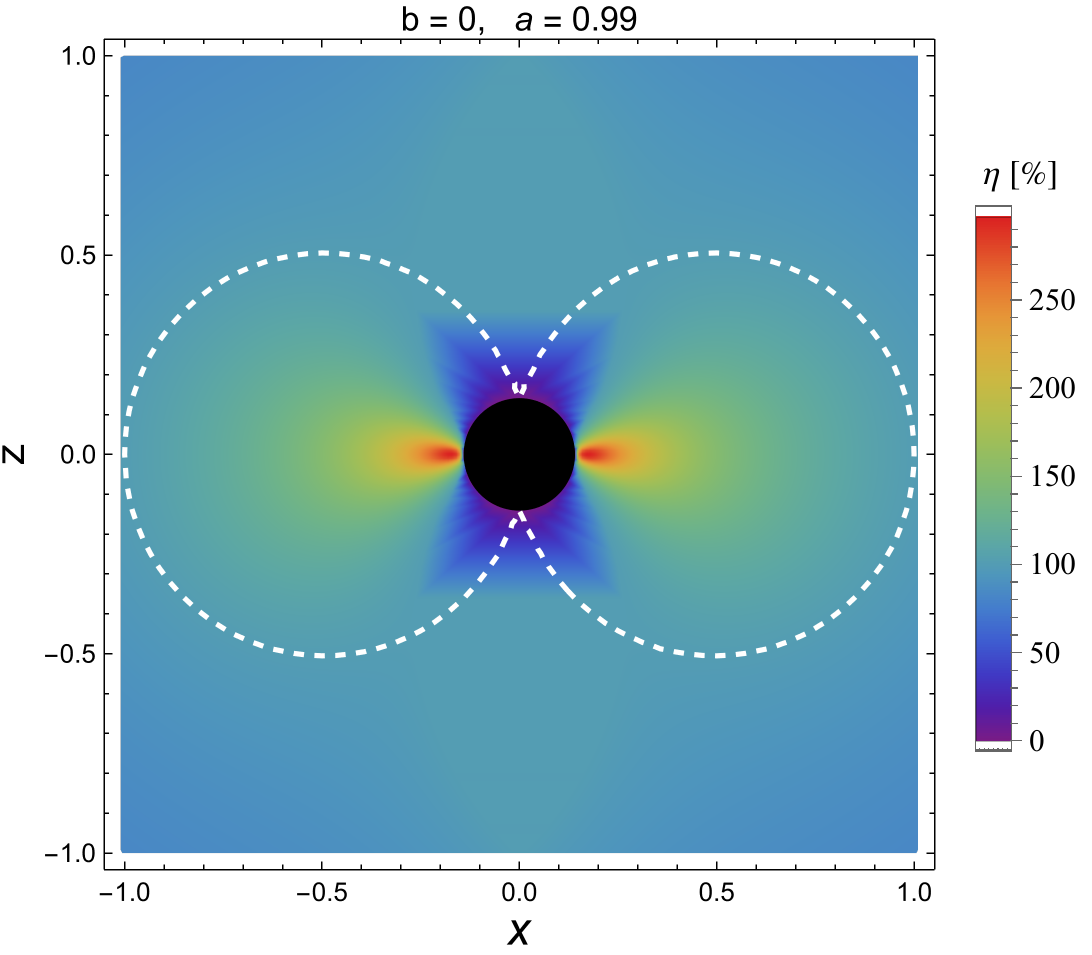}
    \caption{\label{fig:efficiencyeta2} {The efficiency distribution of energy extraction via the MR mechanism from a five-dimensional Kerr BH is plotted with outer horizons (black disks) and ergosphere (white dashed lines) around the five-dimensional BH for two- (left) and single-rotation (right) cases. }}
\end{figure*}

By focusing on the requirements for the extraction of energy by MR, we analyze the phase-space regions under energy extraction conditions $\epsilon_{-}^{\infty}<0$ and $\Delta\epsilon_{+}^{\infty}>0$ by examining solutions of Eq.~(\ref{Eq:essential}). In Fig.~\ref{Fig:energyregion}, the importance of the magnetization parameter and the magnetic field orientation angle can be observed in the phase-space areas where energy extraction could occur. In the left panel, the energetic surfaces (green for accelerated plasma and red to blue for decelerated plasma) are shown for various values of the magnetization parameter from $\sigma_0$=3 to $\sigma_0$=100 when orientation angle is fixed to $\xi=\pi/20$. It can be noticed by the graph that if the magnetization parameter of the plasma is more, phase-space regions which define energy extraction viability through the MR widens to higher r/$\mu$ values and smaller values of rotation parameters $(a, b)$ . Thus it is clear that greater values of magnetization parameter lead to more effective energy extraction through MR. 

The right panel of Fig.~\ref{Fig:energyregion} demonstrates the regions(green for accelerated plasma and orange to light violet for decelerated one) for different orientation angle values from $\xi$=$\pi/20$ to $\xi$=$\pi/6$ when magnetization parameter equals $\xi=100$. It is obvious that when the orientation angle decreases the regions which refers to energy extraction via reconnection broaden to larger quantities of r/$\mu$ and smaller rotation parameters ($a, b$). This comes from the fact that the azimuthal part of the outflow velocity contributes to the extraction of BH rotational energy solely. Though, as we stated earlier, it is much more convenient to take orientation angle as $\xi<\pi/6$ for more efficient energy extraction. Red and blue dashed lines, as seen in Fig.~\ref{Fig:energyregion}, represent outer boundary of ergosphere and external limit of the event horizon, respectively. Hence, the energy extraction can only be driven out within the region bounded by the static surface and the event horizon; i.e., the energy extraction can be more efficient via the MR mechanism in all regions lying inside these bounded surfaces. 

In the next, we analyze the power and energy extraction of MR mechanism for five-dimensional Kerr BH.

\section{\label{Sec:MP-Power-MR} THE EXTRACTED POWER AND EFFICIENCY OF MAGNETIC RECONNECTION}

In this section, we try to define the power that can be extracted from a five-dimensional Kerr BH and the efficiency of MR as a mechanism for energy extraction by applying the Comisso-Asenjo process \cite{Comisso21}. According to the assumptions and analyzes in the previous section, it is obvious that the power which can be taken from the BH by MR is strongly dependent on the decelerated plasma energy at infinity per unit enthalpy. As a result of plasma escape, the power of BH energy which can be extracted via MR is written as 
\begin{eqnarray}
    P_{MR}=-\epsilon_{-}^{\infty}{\mathit{w}}_0 A_{in} U_{in}\, ,
\end{eqnarray}
as is well known, $\epsilon_{-}^{\infty}$ denotes the energy per unit enthalpy at infinity of the infalling plasma, $\mathit{w}_0$ is enthalpy density,  $U_{in}$ is dependent on the regime of the MR mechanism, and $A_{in}$ refers to the area of the inflow cross-section. For simplicity, we shall test the mechanism as in the non-collisional regime, therefore we choose $U_{in}={\cal O}(10^{-1})$, in contrast, $U_{in}={\cal O}(10^{-2})$ is applied for collisional one. The cross-sectional area is defined as $A_{in}$$\sim$$(r_{st}^2-r_{H}^2)$ for rotating BHs in their external cases. In our considerations, we set $a = b \to 0.5$.

In Fig.~\ref{fig:efficiency}, we demonstrate the relationship between the extracted power by MR per unit enthalpy density, $P_{MR}/\mathit{w}_0$, and the dominant reconnection site location, r/$\mu$, for rotation parameters $a=b=0.499$ in the collisionless plasma regime. In the left panel, we can see the relation for various quantities of magnetic field orientating angle from $\xi=0$ to $\xi=\pi/4$ when the magnetization parameter is $\sigma_0=10^4$. Additionally, the right panel represents similar behavior for different values of the magnetization parameter from $\sigma_0=10$ to $\sigma_0=10^5$ and the fixed orientation angle $\xi=\pi/12$. From the plots, we observe a monotonical increase in the power extracted from the five-dimensional Kerr BH as the plasma magnetization increases and the orientation angle decreases. The extracted power reaches a maximum as the location, where the MR occurs, approaches the limiting circular orbit, and subsequently decreases to nearly zero.

We now turn to a key aspect of energy extraction from five-dimensional Kerr BH through the MR mechanism: the efficiency of the process. To assess its viability, it is essential to analyze the efficiency of energy extraction. As we mentioned above, the accelerated and decelerated plasma energy values play an essential role in the investigation of the energetic properties of the model. Therefore, we need a comparison of extracted energy of the plasma, which is considered accelerated one's energy per enthalpy at infinity and total energy: addition of the energies of accelerated and decelerated one as
\begin{eqnarray}
    \eta=\frac{\epsilon_{+}^{\infty}}{\epsilon_{+}^{\infty}+\epsilon_{-}^{\infty}}\, .
\end{eqnarray}
For energy extraction to be available, the condition $\eta>1$ should be satisfied. In Fig.~\ref{fig:efficiencyeta1}, we demonstrate the efficiency of the energy extraction from the five-dimensional Kerr BH as a function of the location $r/\mu$ for single-and two-rotation configurations in the case in which $\sigma=100$ and $\xi=\pi/20$. We note that, in the left panel of Fig.~\ref{fig:efficiencyeta1}, two rotation parameters are equal and nearly maximum values are taken. In contrast, the right panel shows similar behavior for a single rotation case. We can clearly see how eliminating one rotation can affect energy extraction efficiency via the MR mechanism in five-dimensional rotating BH. As can be seen in Fig.~\ref{fig:efficiencyeta1}, the energy efficiency for a single rotation exceeds 250$\%$ and is significantly higher than that for two rotations, as well as for the four-dimensional Kerr BH scenario. This is the remarkable difference
between the single-and two rotation cases of the five-dimensional Kerr BH. For being more informative, in Fig.~\ref{fig:efficiencyeta2}, we also illustrate the efficiency distribution of the MR mechanism at various locations around the five-dimensional Kerr BH with single-and two-rotation configurations. It is clearly seen in Fig.~\ref{fig:efficiencyeta2} that the efficiency gains its maximum value when the reconnection occurs very close to the horizon in the equatorial plane around the five-dimensional Kerr BH.   
\begin{figure*}
    \centering
    \includegraphics[scale=0.68]{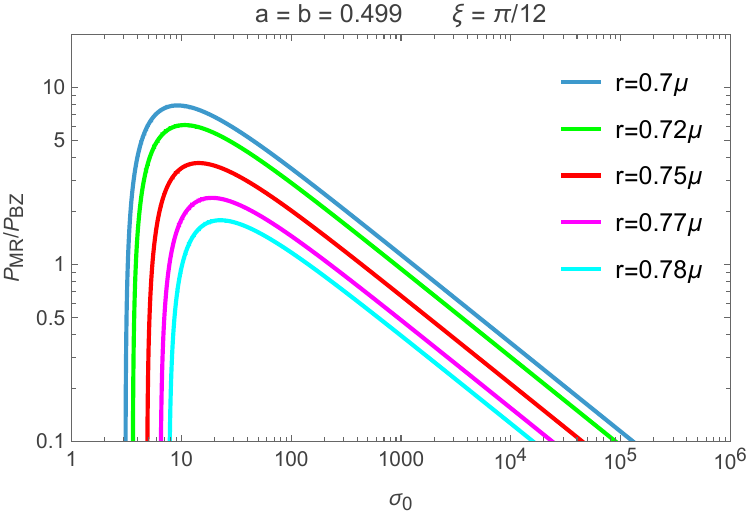}
    \includegraphics[scale=0.83]{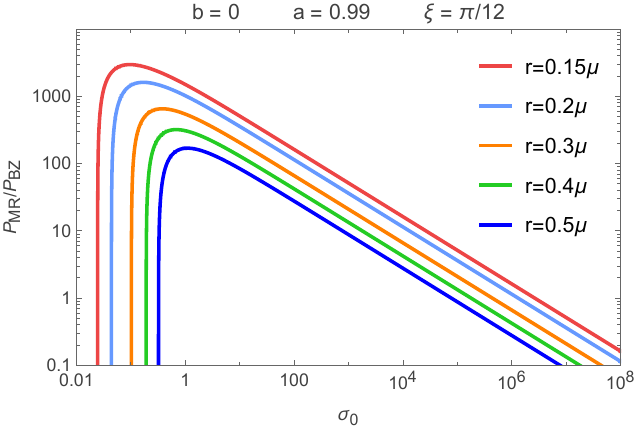}
    \caption{\label{fig:efficiencyratio} {A comparison of two energy extraction mechanisms showing their extracted powers as a function of magnetization parameter for different reconnection point locations. The orientation angle is fixed $\xi=\pi/12$, and single-and two-rotation configurations are considered: $a=b=0.499$ (left panel) and a single rotation case with $b=0, a=0.99$ (right panel). Here, the regime is assumed to be collisionless} }
\end{figure*}

There are several different ways of the energy extraction from BHs. Similarly, the BZ mechanism \cite{Blandford1977} also utilizes the electromagnetic process to extract the rotational energy of a rotating BH by using magnetic field lines and frame-dragging effects. For our analysis, the power which can be taken out from the BH via the BZ process can be expressed as~\cite{Tchekhovskoy10ApJ}
\begin{eqnarray}
   P_{BZ} = \kappa \Phi_{BH}^2 \Omega_{H}^2(1+\chi_1\Omega_{H}^2+\chi_2\Omega_{H}^4+{\cal O}({\Omega_{H}^6}))\, ,
\end{eqnarray}
where $\Phi_{BH}$ refers to magnetic flux which spreads among the ergosphere, $\Omega_{H}$ is the angular velocity at the horizon and $\kappa$, $\chi_1$, $\chi_2$ are fundamental constants, which are chosen by near BH geometries dependent on magnetic field properties. For split monopole geometry of the magnetic field, we take $\kappa=0.05$, in addition, $\chi_1=1.38$ and $\chi_2=-9.2$. We can then write the expression of magnetic flux as $\Phi_{\text{BH}} = \frac{1}{2} \int_{\theta} \int_{\phi} |B^r| \, dA_{\theta\phi}$ and for simplicities we assume $\Phi_{\text{BH}}\sim B_0 \sin \xi \, r_H^2$. 
Then, the comparison of $P_{MR}/P_{BZ}$ can be defined by
 \begin{eqnarray}
\frac{P_{MR}}{P_{BZ}} = \frac{-\epsilon_{-}^{\infty} A_{in} U_{in}}{\kappa \Omega_{H}^2r_{H}^4\sigma_0\sin^2{\xi}(1+\chi_1\Omega_{H}^2+\chi_2\Omega_{H}^4)}\, .
\end{eqnarray}

In Fig.~\ref{fig:efficiencyratio}, we illustrate the ratio of extracted powers,  $P_{MR}/P_{BZ}$, comparing two energy extraction methods: the non-collisional MR and the BZ mechanisms process. The ration is analyzed as a function of the magnetization parameter $\sigma_0$, considering several key quantities at the location point $r/\mu$ where the reconnection occurs. As can be seen in Fig.~\ref{fig:efficiencyratio}, we consider single-and two-rotation configurations for the five-dimensional Kerr BH.  From the left panel, the ratio of maximum extracted power increases as the reconnection location approaches the BH horizon in the two-rotation case, leading to regimes $P_{MR}/P_{BZ}>1$. This indicates that the MR mechanism becomes more efficient than the BZ process in these regions. Unlike two-rotation configuration, the ratio of maximum extracted power becomes significantly high as the reconnection location moves inward, with $P_{MR}/P_{BZ}\gg1$, indicating that MR is significantly more efficient in the single rotation case, as seen in the right panel of Fig.~\ref{fig:efficiencyratio}.

\section{Discussion and conclusion}
\label{Sec:con}

It is widely accepted that rotating astrophysical BHs are among the primary engines powering highly energetic phenomena. Understanding BH energetics is therefore essential for probing the extraction of rotational energy. In this context, higher-dimensional BHs can provide a useful theoretical framework for exploring strong-gravity regimes and high-energy processes, as well as the role of spacetime geometry. In particular, the five-dimensional Kerr BH can serve as an important theoretical laboratory for investigating BH energetics. Unlike its four-dimensional counterpart, a five-dimensional Kerr BH with two independent rotation parameters exhibits a richer ergosphere structure and importantly allows for additional channels of energy extraction. Studying such systems helps us understand whether higher-dimensional gravity can explain high-energy phenomena and allow more efficient energy extraction. Motivated by this, in this work, we extensively investigated the efficiency of energy extraction via the Comisso–Asenjo MR mechanism for the five-dimensional Kerr BH, considering single- and two-rotation configurations.

We studied this MR mechanism in the ergoregion, bounded by the static surface and the event horizon, where energy extraction occurs through bulk plasma consisting of accelerated $\epsilon_{+}^{\infty}>0$ and decelerated $\epsilon_{-}^{\infty}<0$ components. The energy can be driven out from the BH only if the ingoing plasma carries negative energy at infinity per unit enthalpy, while the outgoing plasma extracts positive rotational energy. It was shown that both the plasma energies (i.e., $\epsilon_{+}^{\infty}>0$ and $\epsilon_{-}^{\infty}<0$) and the efficiency of the MR mechanism are strongly influenced by the rotation parameters, magnetization parameter, magnetic field orientation, and the reconnection location. Furthermore, the dependence of the accelerated and decelerated plasma energies indicated that increasing magnetization enhances the efficiency of energy extraction via the MR mechanism. However, this is not the case for the orientation angle of the magnetic field as the smaller angles lead to higher energy extraction efficiency and extracted power from the five-dimensional Kerr BH. 

We further examined the efficiency of energy extraction via the MR process in the five-dimensional Kerr BH with single- and two-rotation configurations. We found that the extraction efficiency strongly depends on the rotation structure in the five dimensions. In particular, the efficiency in the single-rotation case is significantly higher than in the two-rotation case, as well as in the four-dimensional Kerr scenario. The maximum efficiency reaches its maximum value when the reconnection occurs very close to the horizon. This highlights a fundamental distinction between the single-and two rotation configurations in the five dimensions. 

Furthermore, we compared two energy extraction mechanisms, the MR and the BZ mechanisms, by analyzing their extracted powers for the five-dimensional rotating BH with single- and two-rotation configurations. The BZ process extracts rotational energy through electromagnetic fields, driven by the twisting of magnetic field lines due to frame dragging. Our results show that, in the two-rotation case, the ratio of maximum extracted power increases as the reconnection location approaches the horizon, leading to regimes where $P_{MR}/P_{BZ}>1$. This indicates that the MR mechanism becomes more efficient than the BZ process in these regions. In contrast, for the single-rotation configuration, the ratio increases much more rapidly as the reconnection location moves inward, reaching MR $P_{MR}/P_{BZ}\gg1$. This shows that MR is significantly more efficient in the single-rotation case.

Our theoretical studies indicate that the MR process enhances energy extraction in rapidly rotating five-dimensional BHs, making them surprisingly more efficient than their four-dimensional counterparts. They also suggest that five-dimensional BHs are promising sources for explaining high-energy astrophysical phenomena.

\appendix

\bibliographystyle{apsrev4-1}  
\bibliography{gravreferences,Ref_BS1,Ref_BS2}

\end{document}